\documentclass{webofc}
\usepackage[varg]{txfonts}   
\usepackage{color}
\usepackage[usenames, dvipsnames]{xcolor}
\definecolor{bleuuniv}{RGB}{0,54,103}
\definecolor{vertuniv}{RGB}{212,215,0}
\definecolor{mycolor}{RGB}{0,102,204}

\usepackage[load-configurations = abbreviations,alsoload=hep]{siunitx}
\DeclareSIUnit{\sample}{S}
\usepackage{subfig}
\usepackage{tikz}
	\usetikzlibrary{patterns,arrows,decorations.pathmorphing}
\usepackage{caption}
\begin{document}
\title{Multi-wavelength observation of cosmic-ray air-showers with CODALEMA/EXTASIS}
%
%

\author{\firstname{Antony} \lastname{Escudie}\inst{1}\fnsep\thanks{\email{antony.escudie@subatech.in2p3.fr}} \and
        \firstname{Didier} \lastname{Charrier}\inst{1,2} \and \firstname{Richard} \lastname{Dallier}\inst{1,2} \and
        \firstname{Daniel} \lastname{Garc\'{\i}a-Fern\'{a}ndez}\inst{1} \and \firstname{Alain} \lastname{Lecacheux}\inst{3}
        \and \firstname{Lilian} \lastname{Martin}\inst{1,2} \and \firstname{Benoît} \lastname{Revenu}\inst{1,2}
}

\institute{SUBATECH, Institut Mines-Telecom Atlantique -- CNRS/IN2P3 -- Universit\'e de Nantes, Nantes, France 
\and
           Unit\'e Scientifique de Nan\c cay, Observatoire de Paris, CNRS, PSL, UO/OSUC, Nan\c cay, France 
\and
           CNRS-Observatoire de Paris, Meudon, France
          }

\abstract{%
Since 2003, significant efforts have been devoted to the understanding of the radio emission of extensive air shower in the range [20-200] MHz. Despite some studies led until the early nineties, the [1-10] MHz band has remained unused for 20 years. However, it has been measured by some pioneering experiments that extensive air shower emit a strong electric field in this band and that there is evidence of a large increase in the amplitude of the radio pulse at lower frequencies. The EXTASIS experiment, located within the Nançay Radioastronomy Observatory and supported by the CODALEMA experiment, aims to reinvestigate the [1-10] MHz band, and especially to study the so-called "Sudden Death" contribution, the expected electric field emitted by shower front when hitting the ground level. Currently, EXTASIS has confirmed some results obtained by the pioneering experiments, and tends to bring explanations to the other ones, for instance the role of the underlying atmospheric electric field.
Moreover, CODALEMA has demonstrated that in the most commonly used frequency band ([20-80] MHz) the electric field profile of EAS can be well sampled, and contains all the information needed for the reconstruction of EAS: an automatic comparison between the SELFAS3 simulations and data has been developed, allowing us to reconstruct in an almost real time the primary cosmic ray characteristics. 
}
\maketitle
\vspace{-1\baselineskip}
\section{Introduction}
\label{intro}
It is a well known fact that a coherent radio emission is emitted during the development of air shower through the transverse charged current variation induced by the geomagnetic field, and the charge excess mechanism \cite{Kahn206,Askaryan,2015APh....69...50B}. The resulting electric field can be detected by large bandwidth antennas and fast acquisition systems. Commonly, the observations are carried out in the \SI{[20-80]}{\mega\hertz} range (noted MF in the following, for Medium Frequencies) by experiments such as AERA \cite{ThePierreAuger:2015rma} or LOFAR \cite{2013AA...556A...2V} for instance. However, thanks to its design, CODALEMA operates in the \SI{[20-200]}{\mega\hertz} band, referred as EMF in the following for Extended Medium Frequencies. Moreover, the EXTASIS experiment \cite{AntonyICRC2017,EXTASIS}, dedicated to resume the LF (\SI{[1-6]}{\mega\hertz}) study made in the 70's and up to the 90's and hosted by CODALEMA, permits to widen the frequency range, and to demonstrate for the first time that the study of atmospheric showers is possible from 1 to \SI{200}{\mega\hertz} in a continuous way, making it possible to scan different zones and processes of their development and thus offering a more complete and rich view. The characteristics of the primary cosmic rays can be estimated by comparing the electric field detected at ground level to the simulated ones, as described in the following.
\vspace{-1\baselineskip}
\section{Instrumental setup}
\label{Setup}
\begin{figure}[h!]
\begin{center}
  \includegraphics[width=0.5\textwidth]{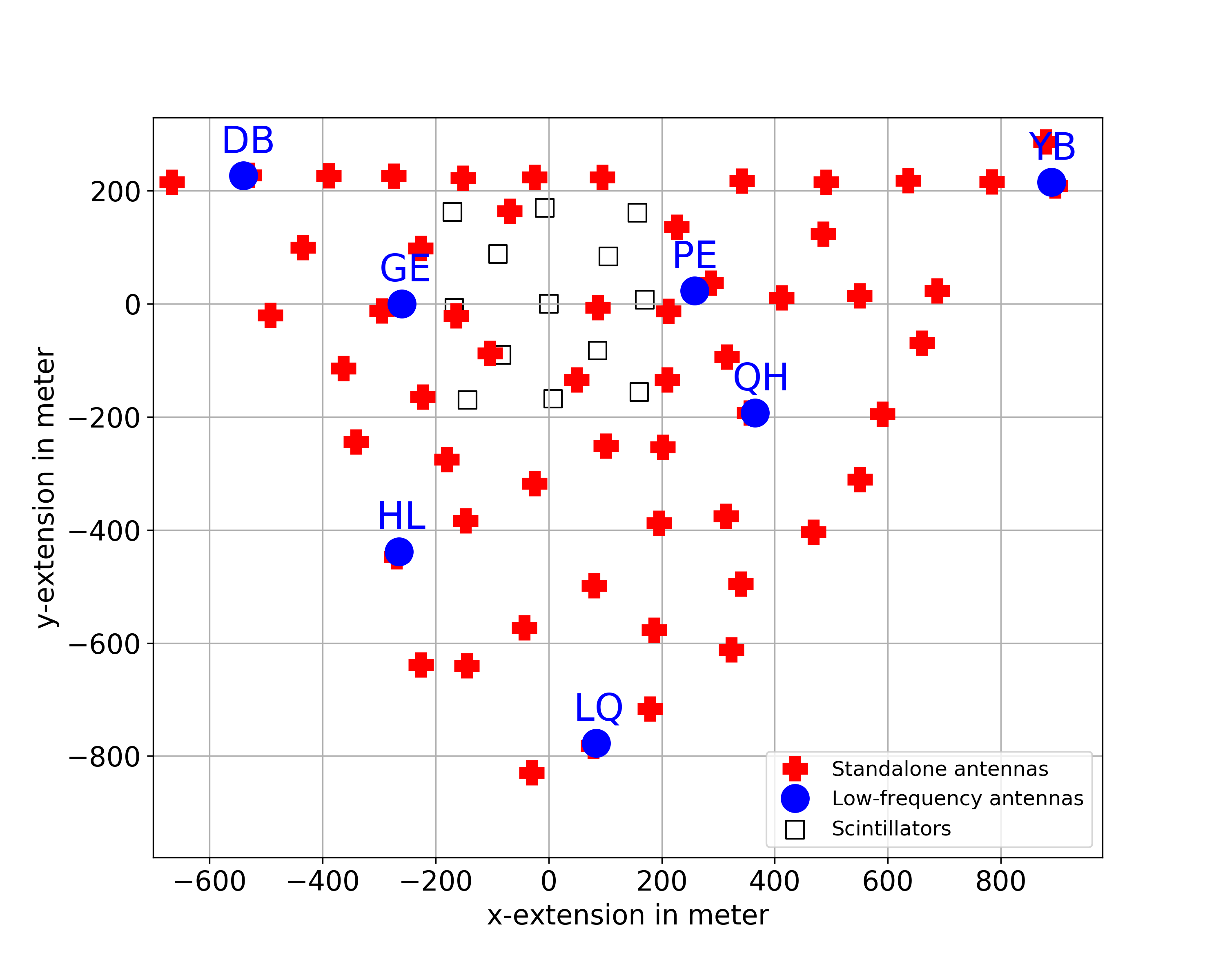}
\caption{Area experiment at the Nançay Radioastronomy Observatory \cite{EXTASIS}. Red crosses represent the 57 standalone antennas of CODALEMA, black squares the 13 scintillators, the blue points represent the 7 LF antennas.}
\label{nancay}
\end{center}
\end{figure}

CODALEMA is hosted since 2002 by the Nançay Radioastronomy Observatory. It is one of the pioneering experiments that have participated in the rebirth of radio detection of cosmic rays. Over the years, the experiment has seen the development of a large collection of detectors, intended to study the properties of the radio emission associated with cosmic ray induced air showers in the energy range from $10^{16}$ to \SI{10^{18}}{\electronvolt}. In its current version, CODALEMA consists essentially of:
\begin{itemize}
\item{a square array (0.4 $\times$ \SI{0.4}{\kilo\metre\squared}) of 13 particle scintillator counters (surface detector),}
\item{a set of 57 so-called ``autonomous'' crossed dipoles and synchronized by GPS dating, operating in the EMF band, distributed over \SI{1}{\kilo\metre\squared},}
\item{a so-called ``Compact Array'' of 10 cross-polarized antennas, arranged in a star shape of \SI{150}{\metre} extension and whose signal acquisition (in the MF band) is triggered by the particle detector.}
\end{itemize}
CODALEMA is today the supporting experiment of EXTASIS, an array of 7 low-frequency antennas also triggered by the particle detector. The LF antenna locations have been chosen to cover the overall Nançay Radioastronomy Observatory area and to have a nearby MF autonomous station. LF atennas are named DB, YB, GE, PE, HL, QH and LQ. Figure~\ref{nancay} shows the experimental area at Nan\c cay (the compact array is not represented).
\vspace{-3\baselineskip}
\section{Analysis method for data/simulations comparisons}
\label{anameth}
\begin{figure}[h!]
\begin{center}
\subfloat{
    \begin{tikzpicture}
    \node[anchor=south west,inner sep=0] (image) at (0,0) {\includegraphics[trim={0 0 1.4cm 1.75cm},clip,width=0.45\textwidth]{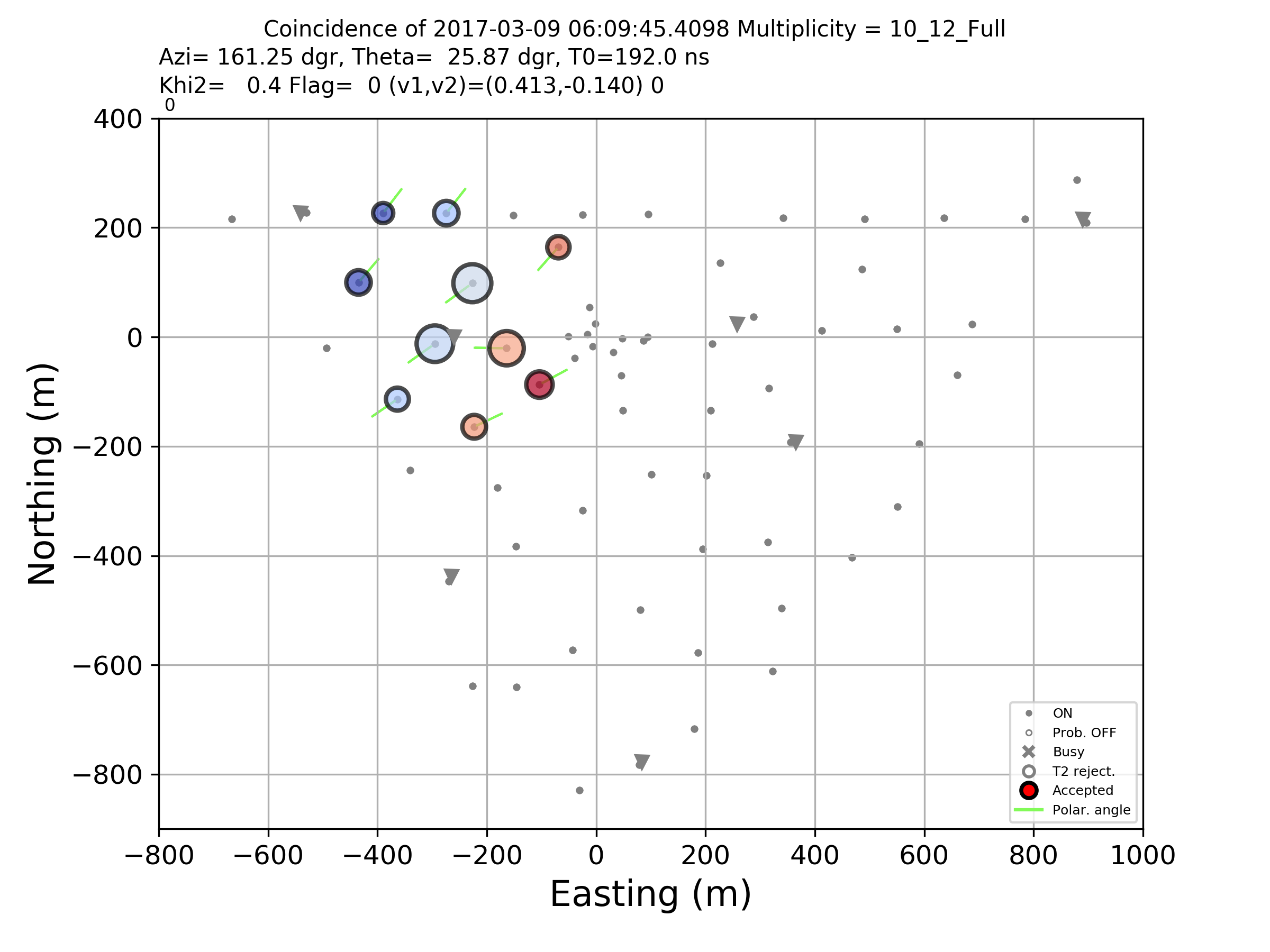}};
    \begin{scope}[x={(image.south east)},y={(image.north west)}]
    \begin{scope}[x={(image.south east)},y={(image.north west)}]
    \end{scope}
    \end{scope}
    \end{tikzpicture}
}
\subfloat{
  \includegraphics[width=0.42\textwidth]{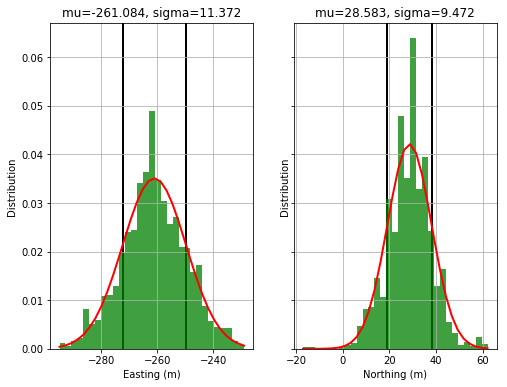} 
}
\\
\subfloat{
  \includegraphics[width=0.45\textwidth]{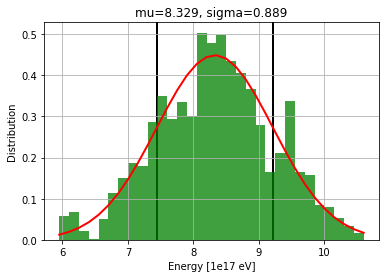}
}
\subfloat{
    \begin{tikzpicture}
    \node[anchor=south west,inner sep=0] (image) at (0,0) {\includegraphics[width=0.42\textwidth]{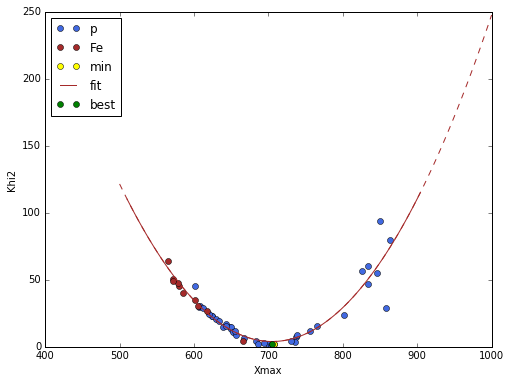}};
    \begin{scope}[x={(image.south east)},y={(image.north west)}]
        \draw[dashed] (0.535,0.1) -- (0.535,0.55);
        \draw (0.535,0.6) node[align=left]{\scriptsize $X_{max} = $\SI{705\pm19}{\gram\per\centi\metre\squared}};
    \end{scope}
        \end{tikzpicture}
}
\caption{Reconstruction of an event detected by CODALEMA. The ground footprint is on the top left figure. The involved standalone antennas are represented by coloured circles, whose colour indicates the timing order in which the signal has been seen by the antennas (from blue, earliest, to red, latest) and size reflects the relative amplitude of the signal (linear scale). The other plots indicate the accuracy on the shower parameters. The red curve is a Gaussian fits of the distributions, the black vertical lines represent the 1$\sigma$ confidence level. The distributions are obtained by propagating the uncertainties on the measured values of the electric field. See text for more details. \vspace{-1.5\baselineskip}}
\label{recoevt}
\end{center}
\end{figure}

For each detected shower, we reconstruct the direction of arrival (DoA) using a plane fit, and using SELFAS (simulation code of the radio signal of cosmic-ray-induced air showers) in its third version (use of a fine description of the atmosphere and improvements to take into account the near field effects) \cite{GATE201838, Garcia-Fernandez:2017yss}, a set of simulations is produced: 40 protons showers and 10 iron showers at an arbitrary energy (\SI{10^{17}}{\electronvolt}) with the corresponding DoA, on a virtual array. Then for different shower core positions, we compare the measured values of the electric field in the MF band to the simulated electric field times a scaling factor to correct from the arbitrary energy. The best agreement is obtained by minimising the chi-squared (see \cite{LilianICRC2017} for more details). To calculate the errors on the estimation of the shower parameters, a propagation of the errors is made. At each step of the comparison, an error on the measured values of the electric field is randomly calculated within the gaussian distribution of the electric field values. This error is added to the electric field values, and the comparison procedure is repeated. After the propagation of the errors, the estimated parameter distributions of the shower is obtained, see figure \ref{recoevt}. Exploiting the capabilities of the CODALEMA instruments, the reconstruction has been improved by: taking into account the high frequency data leading to a reconstruction in the EMF band (very useful for inclined showers); taking into account the information from the Compact Array leading to an hybrid reconstruction. After these improvements, at the end of the analysis chain, the obtained accuracy is \SI{15}{\metre} on the core position, \SI{20}{\%} on the primary energy and \SI{20}{\gram\per\centi\metre\squared} on the atmospheric depth of the maximum of the shower development noted $X_\mathrm{max}$. Figure~\ref{recoevt} shows the reconstruction performances on an event detected by CODALEMA on March, 9$^\text{th}$, 2017.

\section{Example of a multi-wavelengths shower event}
\label{example}

\begin{figure}[h!]
\begin{center}
	\begin{tikzpicture}
	\node[anchor=south west,inner sep=0] (image) at (0,0) {\includegraphics[width=0.5\textwidth]{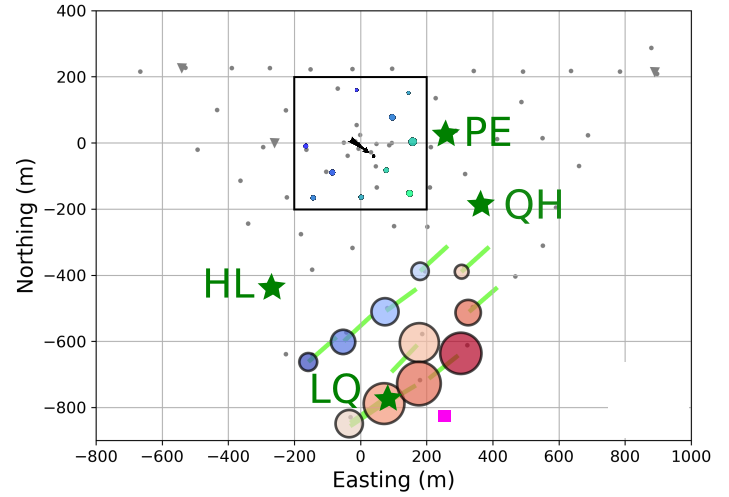}};
	\begin{scope}[x={(image.south east)},y={(image.north west)}]
    \draw[red,line width=1mm,<-] (0.31,0.48) -- ++(canvas polar cs:angle=145,radius=1cm);
	\end{scope}
      \end{tikzpicture}
\caption{Illustrating event seen by the particle detector and some LF and MF antennas \cite{EXTASIS}. The event arrival direction is $\theta=\SI{41\pm2}{\degree}$, $\Phi=\SI{145\pm1}{\degree}$, represented by the red arrow, and its energy \SI{(3.7\pm0.6)\times10^{18}}{\electronvolt}, from the CODALEMA standalone antenna reconstruction. Grey dots represent the standalone antennas, the square area represents the scintillators array region. The stars feature the LF antennas, the involved ones are depicted in green. The involved standalone antennas are represented by coloured circles, whose colour indicates the timing order in which the signal has been seen by the antennas (from blue, earliest, to red, latest), area of circles reflects the relative amplitude of the signal (linear scale). The small green lines close to the circles indicate the orientation of the measured polarization of each MF antenna, nearly orthogonal to the direction of arrival of the event as expected from the dominant geomagnetic mechanism. The estimated shower core location is represented by the magenta square. \vspace{-2\baselineskip}}
\label{mse}
\end{center}
\end{figure}

This section is based on results presented in \cite{EXTASIS}. The ground map of a multi-wavelength observation is shown in figure~\ref{mse}. At the southest part of the CODALEMA array, eleven autonomous stations (circles)  recorded a signal. LF counterparts were registered in four LF antennas. The small green lines close to the circles indicate the orientation of the polarization of each MF antenna, predicted to be orthogonal to the direction of arrival of the event.

\begin{figure}[h!]
\begin{center}
\subfloat[\SI{[30-80]}{\mega\hertz}.]{
  \includegraphics[width=0.435\textwidth]{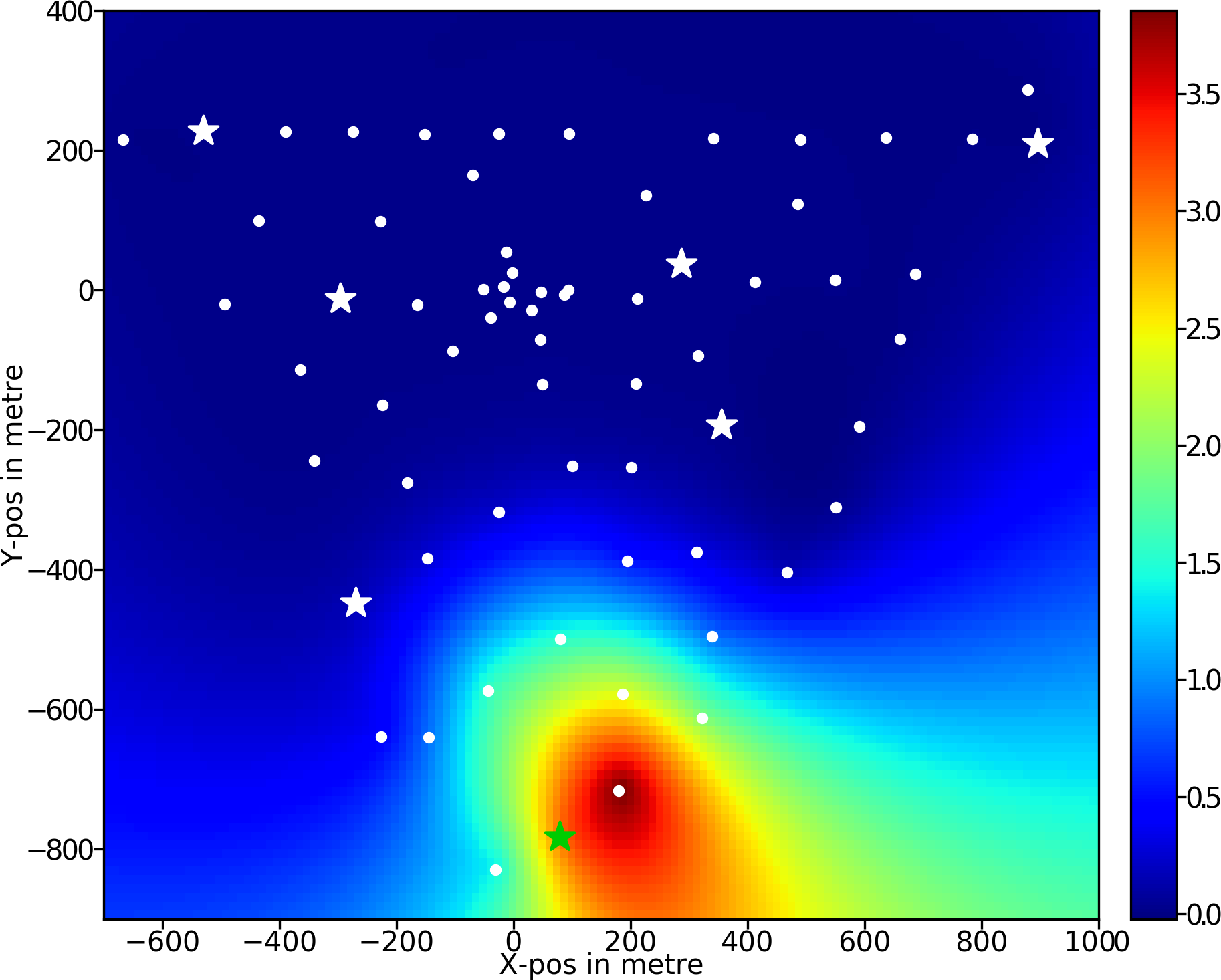}
}
\hspace{10pt}
\subfloat[\SI{[1.7-3.7]}{\mega\hertz}.]{
  \includegraphics[width=0.445\textwidth]{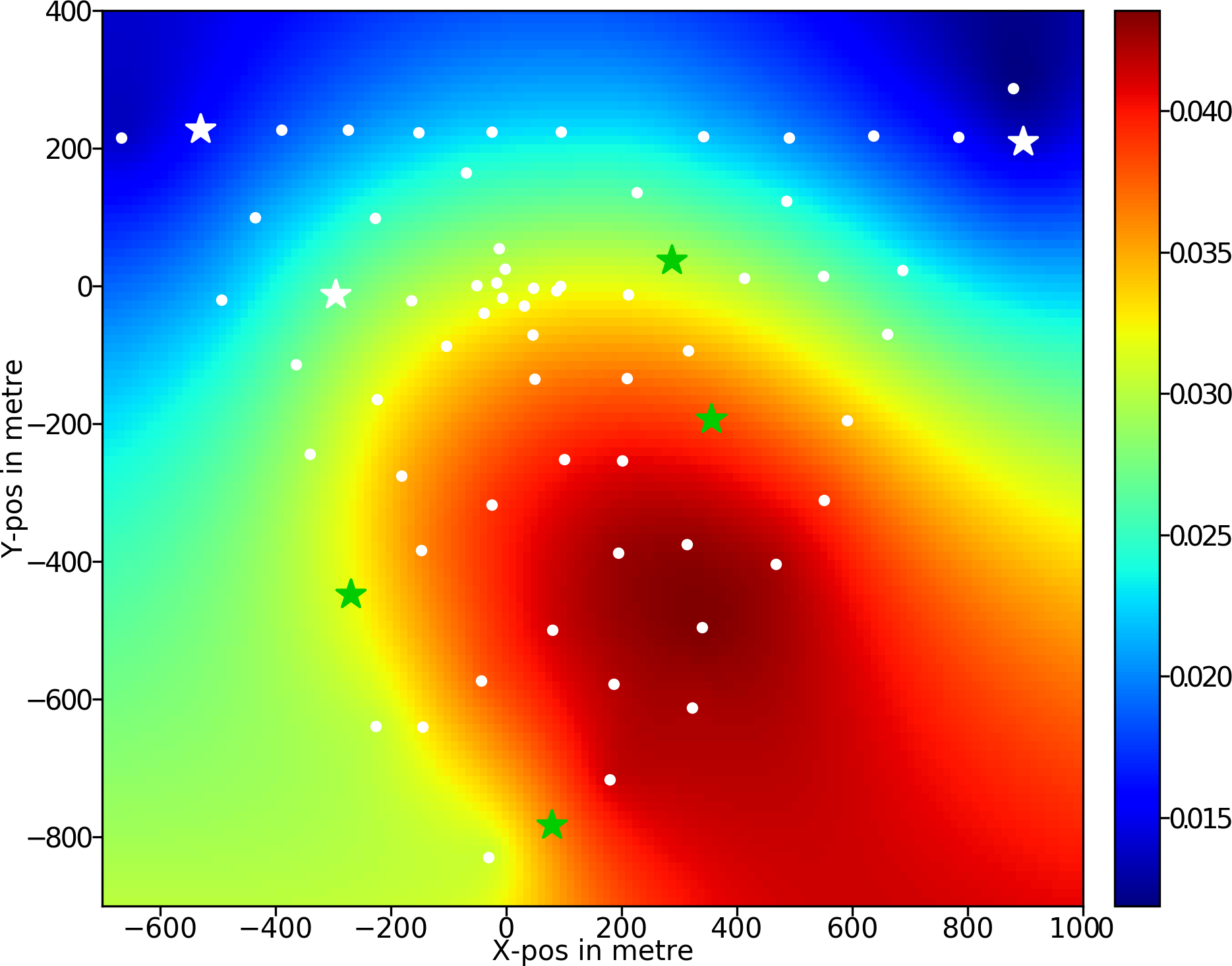}
}
\caption{Ground distribution of the electric field depending on the frequency range predicted by SELFAS3 \cite{EXTASIS}. Left: \SI{[30-80]}{\mega\hertz}. Right: \SI{[1.7-3.7]}{\mega\hertz}. Stars represent the LF antennas, green ones correspond to the involved LF antennas in the event. The LF antenna (PE) located around ($X=\,$\SI{300}{\metre} ; $Y=\,$\SI{20}{\metre}), \SI{850}{\metre} from the shower core location at ground, gives the extent of the detection zone at low frequency. The colour scale, expressed in\SI{}{\milli\volt\per\metre}, is not the same for the two plots: the detected electric field in the LF band is actually two orders of magnitude smaller. \vspace{-2.5\baselineskip}}
\label{ldf}
\end{center}
\end{figure}

\begin{figure}[h!]
\begin{center}
\subfloat{
	\begin{tikzpicture}
	\node[anchor=south west,inner sep=0] (image) at (0,0) {\includegraphics[width=0.49\textwidth]{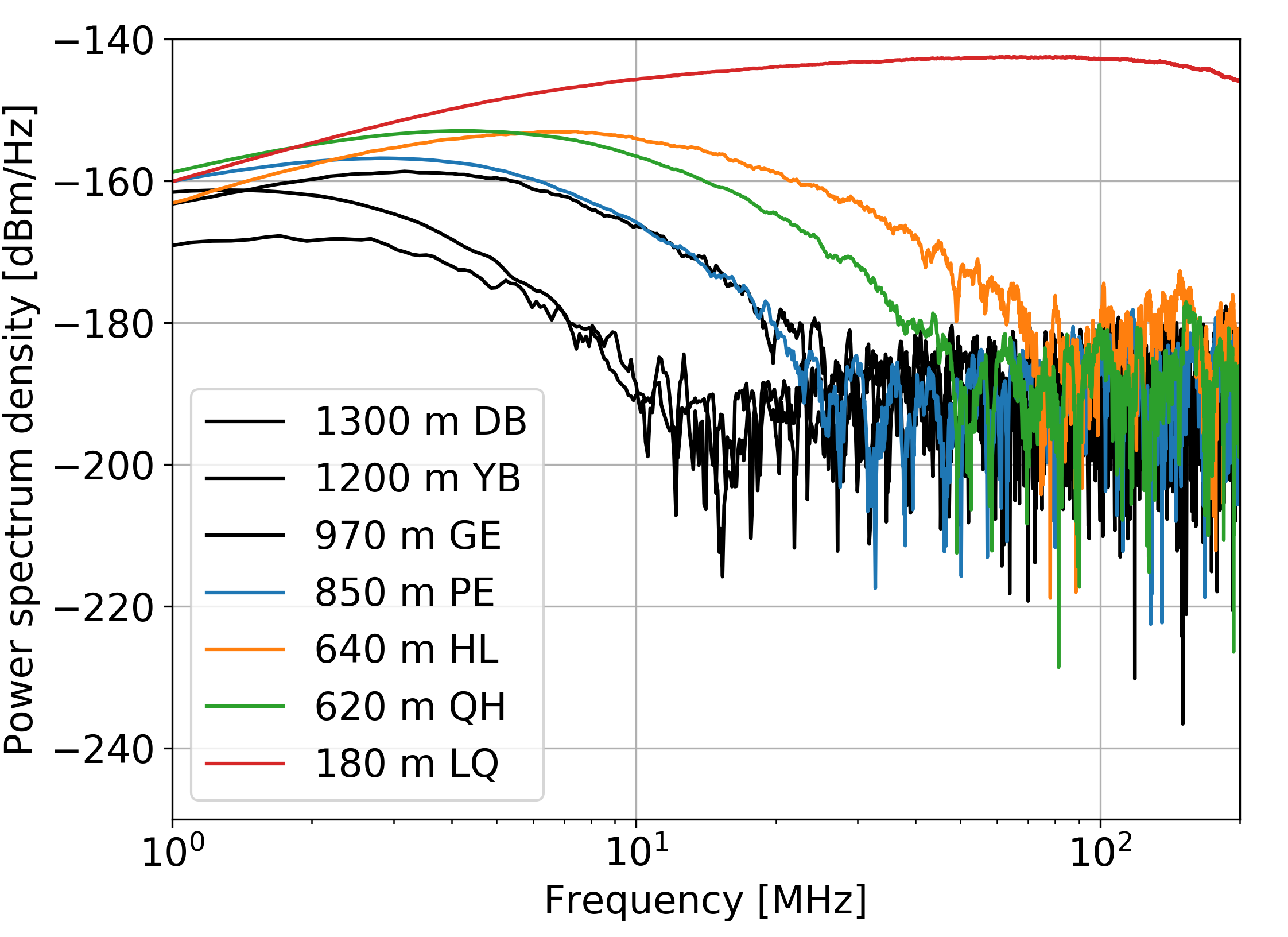}};
	\begin{scope}[x={(image.south east)},y={(image.north west)}]
	\fill[Peach,fill opacity=0.25]
	(0.55,0.96)
	-- (0.9,0.96)
	-- (0.9,0.12)
	-- (0.55,0.12)
	-- cycle;
	
	\fill[LimeGreen,fill opacity=0.25]
	(0.2,0.96)
	-- (0.32,0.96)
	-- (0.32,0.12)
	-- (0.2,0.12)
	-- cycle;

	\end{scope}
      \end{tikzpicture}
}
\subfloat{
  \includegraphics[width=0.497\textwidth]{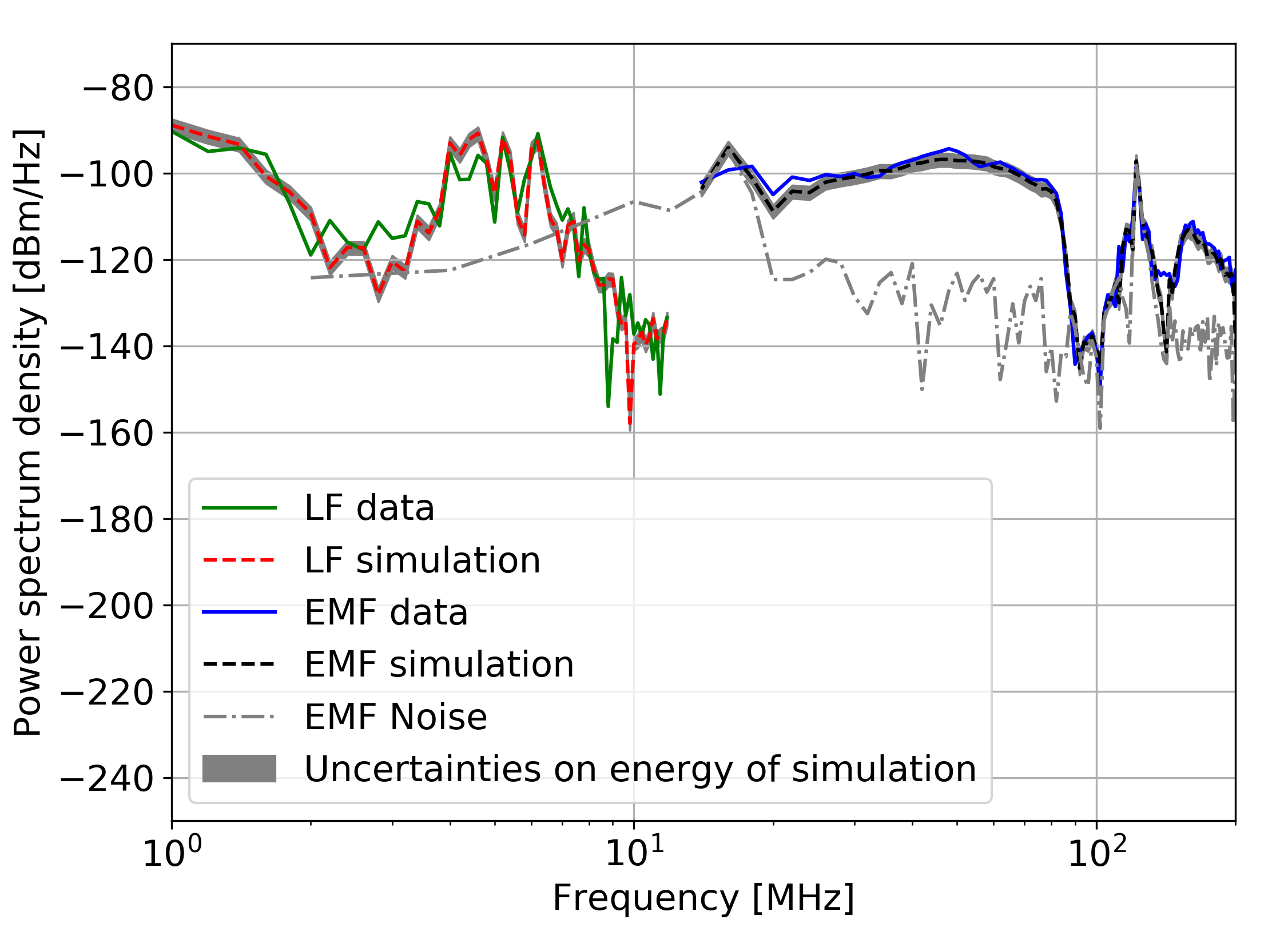}
}
\caption{Left: simulated power spectrum density as a function of frequency calculated at the LF antenna locations, in colour for the involved LF antennas and in black for the others. Distance to shower axis is also indicated. The colored rectangles give the limitations of the LF and EMF bands. Right: convoluted power spectrum density as a function of frequency for the LQ antenna, which detected the air shower. \vspace{-1.5\baselineskip}}
\label{psd}
\end{center}
\end{figure}

The analysis explained in section~\ref{anameth} was applied to this event, giving the best core position at $x=\,$\SI{259\pm35}{\metre} and $y=\,$\SI{-809\pm30}{\metre} ($x=\,$\SI{0}{\metre}, $y=\,$\SI{0}{\metre} being our reference position at the center of the particle detector array), represented by a magenta square in figure~\ref{mse}. We also obtained an estimate of $X_{\mathrm{max}}$ of \SI{715\pm19}{\gram\per\centi\metre\squared} and an energy of \SI{(3.7\pm0.6)\times10^{18}}{\electronvolt}. \\
The interpolated simulated electric fields in \SI{[30-80]}{\mega\hertz} and \SI{[1.7-3.7]}{\mega\hertz} are presented in figure~\ref{ldf}, where we see that the electric field distribution is much wider and flatter at low frequency (right) than at medium frequency (left), with a considerably increased detection range.
Indeed, the LF antenna located around ($X=\,$\SI{300}{\metre} ; $Y=\,$\SI{20}{\metre}) (named PE in figure~\ref{mse}) has detected a signal at \SI{850}{\metre} from the reconstructed shower core location, while the most distant MF antenna is only at \SI{400}{\metre} from the latter. After correction for the antenna equivalent length and acquisition chain gains, we estimate an electric field detection threshold of about \SI{23}{\micro\volt\per\metre} at low frequency, the value detected on the PE antenna. The GE antenna, \SI{\sim100}{\metre} farther from the shower core, has not detected the simulated electric field of \SI{23}{\micro\volt\per\metre} for similar noise conditions. \\
Moreover, the electric field in the LF band is actually smaller than in the EMF band as shown in figure~\ref{ldf} where the colour scale is expressed in\SI{}{\milli\volt\per\metre}, which is in contradiction with the pioneer observations reporting a clear evidence of a strong increase of the radio pulse amplitude when the frequency decreases.
The simulated power spectrum densities (PSD) as a function of frequency, for different LF antenna locations and for different shower axis distances are shown in figure~\ref{psd} (in colour for the involved LF antennas and in black for the others). The PSD quickly drops in the EMF band with the shower axis distance, while it decreases much more slowly in the LF band. This result indicates again that the detection range should be larger in the LF band than in the EMF band, and explain why only the southernmost LF antenna (LQ) presents a signal jointly with the associated standalone antenna.

Moreover, at a given shower axis distance (for the LQ antenna for instance), there are \SI{10}{\deci\bel m\per\hertz} between the maximum in the EMF band and the maximum in the LF band, also indicating that the electric field in the EMF band is larger than in the LF band. For the LQ antenna, the PSD has been convoluted with the antenna response in order to be compared with the raw data, as shown in figure~\ref{psd}-right which presents the PSD of the signal of the shower development over the whole frequency band: LF data are represented by the green line, EMF data by the blue line, and the red and black dashed lines show the convoluted simulated power spectrum density in LF and EMF band respectively, in which we have added the noise of the corresponding band. The noise-added, convoluted simulations are in good agreement with the data, showing a good understanding of our LF and MF instruments, but also a good radio reconstruction of the characteristics of the primary cosmic ray. \smallskip

As mentioned in \cite{EXTASIS}, since the installation of the complete instrumental setup of EXTASIS (March 2017) and until the end of year 2017, only 18 LF events have been detected. This low rate of detection can be explained by harsh atmospheric noise conditions. In the best case, the atmospheric noise level remains 10 times higher than the amplitude of the signal that we want to detect. Moreover, we have found a correlation with the atmospheric electric field, that probably amplifies the transient signal and lowers again the real detection efficiency in normal conditions. For these reasons, the LF band seems not very promising in terms of efficiency for the detection of air showers.

\vspace{-0.5\baselineskip}
\section{Conclusion}
\label{conclusion}

CODALEMA/EXTASIS is a multi-wavelengths experiment observing cosmic-ray induced air-showers in \SI{[1-200]}{\mega\hertz}. Its unique capabilities permit to better constrain the reconstruction of the cosmic ray properties, and to demonstrate for the first time that the study of atmospheric showers is possible over a very wide frequency band in a continuous way, making it possible to scan different zones and processes of the air shower development and thus offering a more complete and rich view. However, due to the low efficiency of detection in the LF band, the most interesting frequency band for detecting air showers remains the \SI{[20-200]}{\mega\hertz} band.

\vspace{-0.5\baselineskip}
\section*{Acknowledgements}
We thank the R\'egion Pays de la Loire for its financial support of the Astroparticle group of Subatech and in particular for its contribution to the EXTASIS experiment, and the PNHE (Programme National Hautes Energies) from the french institutes IN2P3 and INSU for having also always supported the CODALEMA experiment, both financially and scientifically.

\vspace{-1\baselineskip}
%
\bibliographystyle{woc} 
\bibliography{biblio.bib}

\end{document}